\documentclass{aastex} 
 
\usepackage{graphicx}
 
\usepackage{emulateapj5}
\usepackage{times}

\def\fdg{\hbox{$^\circ\!.$}}
\def\lsim{\lower.5ex\hbox{$\; \buildrel < \over \sim \;$}} 
\def\gsim{\lower.5ex\hbox{$\; \buildrel > \over \sim \;$}} 
\def\lax {\ifmmode{_<\atop^{\sim}}\else{${_<\atop^{\sim}}$}\fi} 
\def\gax {\ifmmode{_>\atop^{\sim}}\else{${_>\atop^{\sim}}$}\fi} 
 
\def\gtorder{\mathrel{\raise.3ex\hbox{$>$}\mkern-14mu 
\lower0.6ex\hbox{$\sim$}}} 
\def\ltorder{\mathrel{\raise.3ex\hbox{$<$}\mkern-14mu 
\lower0.6ex\hbox{$\sim$}}}

\makeatletter
\newenvironment{inlinetable}{%
\def\@captype{table}%
\noindent\begin{minipage}{0.999\linewidth}\begin{center}\footnotesize}
{\end{center}\end{minipage}\smallskip}
 
\newenvironment{inlinefigure}{%
\def\@captype{figure}%
\noindent\begin{minipage}{0.999\linewidth}\begin{center}}
{\end{center}\end{minipage}\smallskip}

\long\def\aftertabcaption#1{\begin{minipage}{0.999\linewidth}#1\end{minipage}}

\makeatother

\usepackage{mathptmx}

\begin{document} 

\submitted{Received 2002 February 25; Accepted 2002 April 19}
\journalinfo{The Astrophysical Journal, 571:L000, 2002 June 1}

\title{Can  Any ``Invariants'' Be Revealed in   Quasi-periodic Phenomena 
Observed From Sco~X-1?} 

\author{Sergey Kuznetsov\altaffilmark{1} and Lev Titarchuk \altaffilmark{2,3}}

\altaffiltext{1}{High-Energy Astrophysics Department, Space Research
Institute, Profsoyuznaya 84/32, Moscow, 117997,
Russia; sik@hea.iki.rssi.ru} 
\altaffiltext{2}{George Mason University/CEOSR/NRL;
lev@xip.nrl.navy.mil}
\altaffiltext{3}{NASA Goddard Space Flight Center, code 661, 
Laboratory for High Energy
Astrophysics, Greenbelt MD 20771; lev@lheapop.gsfc.nasa.gov}

\shorttitle{``INVARIANTS'' IN QUASI-PERIODIC PHENOMENA FROM SCO X-1}
\shortauthors{KUZNETSOV \& TITARCHUK}
\received{25 February 2002} 
\accepted{25 April 2002} 

\thispagestyle{empty}

\begin{abstract} 

Using large number of Rossi X-ray Time Explorer observations of Sco
X-1 we present a detailed investigation of the transition layer (TL)
and the relativistic precession (RP) models.  These models predict the
existence of the invariant quantities: an inclination angle $\delta$
of the magnetospheric axis with the normal to the disk for the TLM and
a neutron star (NS) mass $M_{NS}$ for the RPM. Theoretical predictions
of both models are tested and their self-consistency is checked. We
establish that: (i) The inferred $\delta-$value is
$5\fdg56\pm0\fdg09$.  Correlation of the $\delta-$values with the
horizontal branch oscillation (HBO) frequency is rather weak.  (ii)
There is a strong correlation between an inferred $M_{NS}$ and the HBO
frequency in the RPM frameworks.  (iii) We infer $M_{NS}$ for
different assumptions regarding the relations between the HBO
frequency $\nu_{HBO}$ and the nodal frequency $\nu_{nod}$.
We find that the inferred $M_{NS}=(2.7\pm 0.1)M_{\odot}$ cannot be consistent
with any EOS of NS matter. 
We conclude that RPM fails to describe the data while TLM seems to be
compatible.
\end{abstract} 


\section{Introduction} 

Kilohertz quasi-periodic oscillations (QPOs) have been discovered by
the Rossi X-ray Timing Explorer (RXTE) in a number of low mass X-ray
binaries (Strohmayer et al.  1996, van der Klis et al. 1996). The
presence of two observed peaks with frequencies $\nu_1$ and $\nu_2$ in
the upper part of the power spectrum became a natural starting point
in modeling the phenomena.  Attempts have been made to relate $\nu_1$
and $\nu_2$ and the peak separation $\Delta \nu= \nu_2-\nu_1$ with the
neutron star (NS) spin. In the beat frequency model (Miller, Lamb \&
Psaltis 1998) the kHz peak separation $\Delta \nu$ is considered to be
close to the NS spin frequency and thus $\Delta \nu$ is predicted to
be constant. In other words $\Delta\nu$ is an ``invariant'' which does
not vary when the kHz QPOs frequencies change.  However observations
of kHz QPOs in a number of binaries (Sco X-1, 4U 1728-34, 4U 1608-52,
4U 1702-429 and etc) show that the peak separation decreases
systematically when kHz frequencies increases (see a review by van der
Klis 2000, hereafter VDK).  For Sco X-1 VDK found that the
peak separation of kHz QPO frequencies changes from 320 Hz to 220 Hz
when the low peak $\nu_1$ changes from 500 Hz to 850 Hz.  The lower
frequency part of the power spectrum, contains two horizontal branch
oscillation (HBO) frequencies $\nu_{HBO}\sim 45$ Hz and
$\nu_{2HBO}\sim 90$ Hz (presumably the second harmonic of $\nu_{HBO}$)
which slowly increase with increase of $\nu_1$ and $\nu_2$ (van der
Klis et al. 2000). {\it Any consistent model faces a challenging task
of describing the dependences of the peak separation $\Delta\nu$ and
HBO frequency $\nu_{HBO}$ on $\nu_1$ and $\nu_2$} .

\subsection{QPO model description}
 
There are two other QPO models in the literature which infer the
relations between $\nu_1$, $\nu_2$ and $\nu_{HBO}$.

The transition layer model (TLM) was introduced by Titarchuk, Lapidus
\& Muslimov (1998), hereafter TLM98, to explain the dynamical
adjustment of a Keplerian disk to the innermost sub-Keplerian boundary
conditions (e.g. at the NS surface).  TLM98 argued that a shock should
occur where the Keplerian disk adjusts to the sub-Keplerian flow.  It
was suggested by Osherovich \& Titarchuk (1999), hereafter OT99, that
the radial oscillations of the fluid element that bounced from the
disk shock region (presumably at the adjustment radius) would be seen
as two independent oscillations in the radial and the vertical
directions due to the presence of a Coriolis force in the
magnetospheric rotational frame of reference. Simultaneous
measurements of the frequencies of the two kHz QPOs and HBO harmonics
in a wide frequency range allow one to derive the angle $\delta$.
Titarchuk \& Osherovich (2001, hereafter TO01), claim that for the
sources GX 340+0, Sco X-1, 4U 1702-42 and 4U 0614+09 the inferred
angle $\delta$, (see Eq 1 in TO01) stays the same over significant
range of the observed QPO frequencies.
The low branch frequency $\nu_L$, the Kepler frequency $\nu_{\rm K}$
and the hybrid frequency $\nu_h$, as they are introduced by OT99 are
eigenfrequencies of the oscillator.  However these frequencies are
revealed in the observations as the {\it resonance} frequencies
$\nu_{HBO}$, $\nu_1$ and $\nu_2$ which are broadened as a result of
the (radiative) damping in the oscillator (see TLM98, Eq. 15).
Furthermore, the resonance frequencies $\nu_{HBO}$, $\nu_1$, $\nu_2$
are shifted with respect to the eigenfrequencies $\nu_L$, $\nu_{\rm
K}$, $\nu_h$.  The frequency shift and random errors of the
eigenfrequencies depend on the damping rate of oscillations $\lambda$
(see details in \S 2).  One should keep in mind that the systematic
and random errors in the centroid frequency determination due to this
resonance shift can be a factor of a few larger than the statistical
error in the determination of the centroid frequency.  In the present
data analysis (as the first approximation) we assume that $\nu_{\rm
K}=\nu_1$, $\nu_h=\nu_2$ and $\nu_L=\nu_{HBO}$.  We discuss the
results of this analysis in \S 2.


The relativistic precession model (RPM) is related to high speed
particle motion in strong gravitational fields, leading to
oscillations of the particle orbits. Bardeen, Press \& Teukolsky
(1972), Okazaki, Kato \& Fukue (1987), Kato (1990) and later Stella et
al. (1999) studied precession of the particle orbit under influence of
a strong gravity due to the GR effects. In order to generate these
oscillations one should assume that the particle orbit is not
precisely in the equatorial plane of the compact object. The
inclination angle between the particle orbit plane and the equatorial
plane can be infinitesimal. For a nonrotating configuration of the
central body (in the Schwarzchild's treatment) the angular velocity is
calculated using a classical Kepler's formula.  In the general case,
when the intrinsic angular momentum, $a\ne0$ the azimuthal frequency
is expressed in units $G=1$, $c=1$ as follows (Bardeen et al., 1972;
Stella et al., 1999)
\begin{equation}
\nu_{\phi}=(M/r^3)^{1/2}\left\{2\pi[1+a(M/r^3)^{1/2}]\right\}^{-1}.
\end{equation} 
Hereafter, we consider a corotating neutron star and a probe particle,
namely with $\nu_{\phi}>0$ and $a>0$. An epicyclic frequency $\nu_r$
along with $\nu_{\phi}$ determine the orbital periastron rotation
$\nu_{per}=\nu_{\phi}-\nu_r$. A precession frequency of the node line
$\nu_{nod}$ is calculated using $\nu_{\phi}$ and a frequency of
oscillations in the direction which is perpendicular to the orbital
plane, $\nu_{\theta}$, namely $\nu_{nod}=|\nu_{\phi}-\nu_{\theta}|$.
Formulas for $\nu_r$ and $\nu_{\theta}$:
\begin{equation}
\nu_{\theta}=\nu_{\phi}[1-4a(M/r^3)^{1/2}+3a^2/r^2]^{1/2},
\end{equation} 
\begin{equation}
nu_r=\nu_{\phi}[1-6M/r +8a(M/r^3)^{1/2}-3a^2/r^2]^{1/2}
\end{equation} 
have been obtained by Okazaki, Kato, \& Fukue (1987) and Kato (1990).
In the framework of the RPM (e.g. Stella et al., 1999) a frequency
of Keplerian rotation $\nu_{\phi}$ is related to the higher kHz peak
$\nu_2$, and a frequency of the periastron precession $\nu_{per}=
\nu_{\theta}-\nu_r$  
is  related to a lower kHz peak $\nu_2$  
\footnote{In the Kerr's geometry $\nu_{\phi}\neq\nu_{\theta}$
(for the Schwarzchild's case with a dimensionless angular momentum
$a=0$ these frequencies coincide $\nu_{\phi}=\nu_{\theta}$).}.
The HBO frequency $\nu_{HBO}$ is related to the nodal precession 
$\nu_{nod}$.  Stella et al. (1999)
proposed that $\nu_{HBO} $ can be an even harmonic of $\nu_{nod}$.
{\it Thus  equations (1)-(2) allow one to find the neutron star mass 
$M_{NS}$ and $a$ using the observable frequencies $\nu_{HBO}$, $\nu_1$
and $\nu_2$}. 

The goal of this {\it Letter} is to test the invariant predictions of
the TLM and the RPM using extensive RXTE observations of QPO
phenomena.
The {\it Letter} reports the results of the detailed data analysis
from Sco X-1 collected by RXTE during 4 years of observations from
1996 to 1999.  In \S 2, we describe the RXTE data which we use to
construct the power spectra in the frequency range from $\sim0.03$ Hz
to $\sim2050$ Hz.  In \S 2 we also give details of the resonance
effect on the eigenfrequency restoration using the observed QPO
frequencies and present comparisons of the predictions of the QPO models with
the RXTE observations.
Summary and conclusions are drawn in \S 3. 

\section{Observations and Data Analysis}  

We used data of from the RXTE proportional counter array (PCA)
instrument (Jahoda et al. 1996) retrieved from the high energy
astrophysics archive of NASA/GSFC. Sco~X-1 was observed by RXTE during
11 observations (10056, 10057, 10059, 10061, 20053, 20426, 30035,
30036, 30406, 40020, 40706) in 1996--1999. Data were collected from
either energy channels 0-87 ($\sim 2-20$ keV) or 0-249 ($\sim 2-60$
keV) at resolutions of $\sim 244\mu$s or higher. 
For observations when ``double-event'' data\footnote{Two events
detected within 6$\mu$s at two different anodes of detector are due to
charged particles and high count rate from very bright sources like
Sco~X-1. In this case the part of the source flux dominates in
double-events data.} were available, we combined them with single-event
to increase statistics. We constructed the power density spectra (PDS)
in the frequency range from 0.03125~Hz to 2048~Hz to analyze the
Sco~X-1 variability.  The features of the low frequency and high
frequency variability were investigated in the range $0.1-256$ Hz and
$256-2048$ Hz respectively.
The PDS was fit by a constant (due to dead time
effect, see details in Vihlinin  et al. 1994) and a few Lorentzians to
determine QPO frequencies and we added the following term
$P(\nu)=A\nu^{-\alpha}[1+(\nu/\nu_{break})^{\beta}]^{-1},$
to our model to approximate low-frequency noise component.

\begin{inlinefigure}
\centerline{\includegraphics[width=0.999\linewidth]{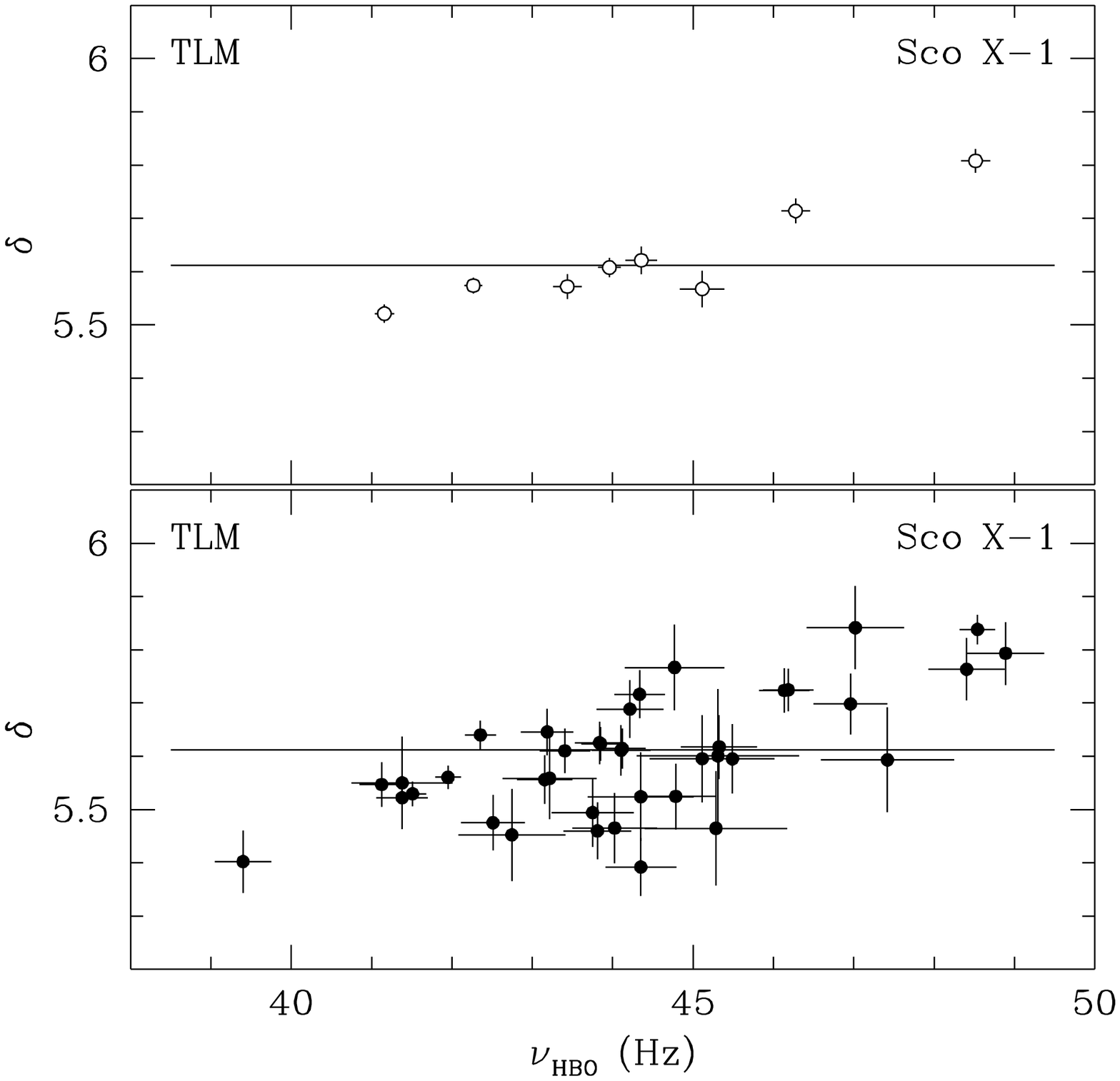}}
\caption{Dependence of the inferred $\delta$ angle on the HBO frequency,
$\nu_{HBO}=\nu_L$. Low panel: $\delta$ for a particular observation
with a duration $\sim$3 ks. Upper panel: that presented in upper panel is
averaged over 5 HBO frequencies. Approximations by a constant value
are shown by solid lines.}
\label{fig1} 
\end{inlinefigure}

\subsection{HBO frequencies and its harmonics} 
Among all analyzed spectra where a kHz QPO pair was detected we
selected only those for which an HBO was detected as well and at a
significance more than 4 $\sigma$. The second harmonic $\nu_{2HBO}$ is
present for nearly all of these $\nu_{HBO}$. It is worth noting that
the significance of the HBO peaks decrease with the frequency which is
a well known effect for quite a few sources noted by VDK.  Because of this, we cannot find HBO harmonics for HBO
frequencies higher than 48 Hz ($\nu_{2HBO}\gax96$ Hz) with reliable
level of significance.  Ratio $r_{2/1}=\nu_{2HBO}/\nu_{HBO}$, on the
average, is slightly below than 2, namely $r_{2/1}=1.965\pm
0.004$. The observational appearance of the HBO harmonics is
presumably a combination frequency effect along with a resonance in
weakly nonlinear oscillation system (Landau \& Lifshitz 1965,
hereafter LL).

\subsection{Resonance effect}
The resonance in weakly nonlinear systems occurs at the
eigenfrequencies of the system $\nu_0$ when the frequency of the
driving force $\gamma$ is $\gamma\approx p\nu_0/q$ and $p$, $q$ are
integers (LL).  The main peak resonance power (for $p=1$ and $q=1$) is
the strongest among all the harmonics because the peak power
diminishes very quickly with the increase of $p$ and $q$ (LL). The
maximum of the main resonance amplitude for the linear and nonlinear
oscillations is not precisely at the eigenfrequency $\nu_0$, but rather
it is shifted to the frequency $\nu_r=\nu_0+\eta$.  For a linear
oscillator $\nu_r$ depends on the damping rate $\lambda$ (see LL and
and TLM98, Eq. 15 for the power of the forced oscillations), i.e.
$\nu_r=(\nu_0^2-2\lambda^2)^{1/2}$ and thus the frequency shift
\begin{equation}
\eta=(\nu_0^2-2\lambda^2)^{1/2}-\nu_0.
\end{equation}
From Eq. (3) it follows that  
$\eta\approx -\lambda^2/\nu_0=-\lambda/Q$  when $Q=\nu_0/\lambda\gg 1$. 
 LL also shows  that the half-width of the resonance curve 
$\Delta\nu_{FW}\approx\lambda$ for $Q\gg 1$.
Taking a differential from the left and  right hand sides of the formula for $\nu_r$  
one can derive a formula for the random error of the eigenfrequency
determination $\Delta \nu_0$ due to the resonance effect:
$\Delta\nu_0=\nu_r\Delta\nu_r/\nu_0+2\lambda\Delta\lambda/\nu_0$.  
Now one can estimate the ratio of the systematic shift and random
error to $\Delta \nu_{QPO}$. They are $\eta/\Delta \nu_{QPO}\approx
\Delta\nu_{FW}/Q\Delta\nu_{QPO}\gax 8-10$ and $\Delta\nu_0/\Delta
\nu_{QPO}> 2\nu_{FW}\Delta\nu_{FW}/(\nu_{QPO}\Delta\nu_{QPO})\gax 2-3$ for
frequencies presented by Jonker et al. (2002) in Tables~2 and 4
\footnote{Jonker et al (2002) presented the analysis of the $\delta-invariant$
of the TLM. They found that $\delta=6^0.3\pm 0\fdg1$ and
$\delta=6\fdg1\pm 0\fdg.2$ for GX 5-1 and GX 17+2 respectively. But
they pointed out the very large $\chi^2_{red}=95.2/9$ and
$\chi^2_{red}=377./9$ for GX 5-1 and GX 17+2 respectively.  Their
$\chi^2$ calculations include the statistical error only but do not
account for the systematic resonance shift $\eta$ and for the random
resonance error $\Delta\nu_0$.}.
\begin{inlinefigure}
\centerline{\includegraphics[width=0.999\linewidth]{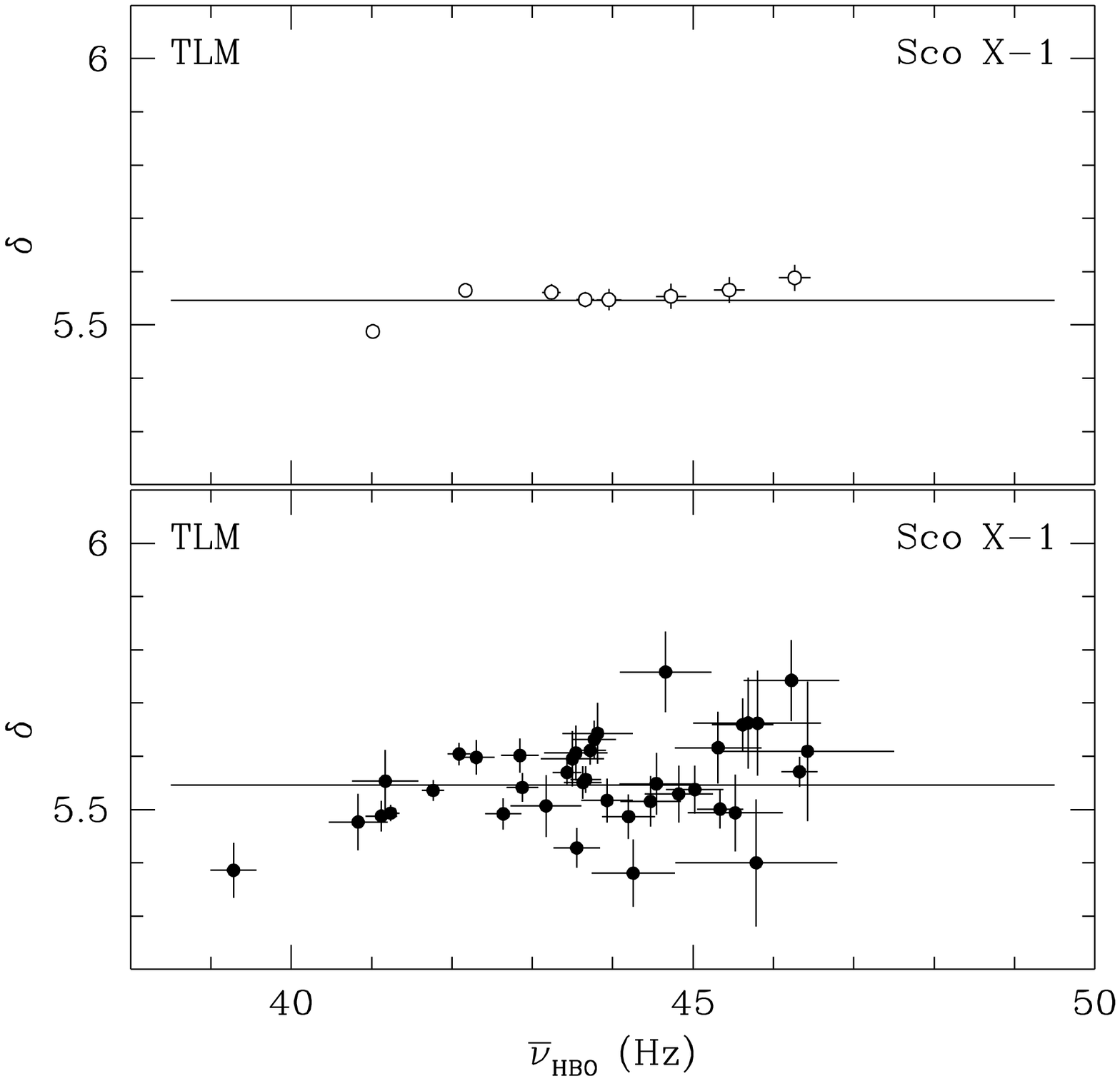}}
\caption{Dependence of the inferred $\delta$ angle on the HBO frequency, 
$\nu_L=\overline{\nu}_{HBO}=\case{1}{2}(\nu_{HBO}+\case{1}{2}\nu_{2HBO})$. Low
panel: $\delta$ for a particular observation with a duration $\sim$3~ks.
Upper panel: that presented in upper panel is averaged over 5 HBO
frequencies. The best fits of the $\delta-$ values are shown by solid
lines.}
\label{fig2}
\end{inlinefigure}
We can conclude that Jonker's et al. $\chi^2$ for the $\delta-$mean
would be significantly reduced if this effect were taken into account.

In the nonlinear case the systematic and random shifts are also
affected by the amplitude of the driving force and the relative
weights (coefficients) of the nonlinear terms (LL). It is not by
chance that our observed values of $r_{2/1}$ differ from 2, {\it since
the apparent observed resonance frequency $\nu_{HBO}$ is shifted with
respect to the eigenfrequency $\nu_0$}. This realization is
particularly important when one compares the observationally
identified frequencies $\nu_{HBO}$, $\nu_1$ and $\nu_2$ and those
predicted by the QPO models.  In order to lessen the uncertainty in
determining the real eigenfrequency value as well as to decrease the
systematic error caused by this nonlinear effect, we use two values
for the low-frequency identification: $\nu_{HBO}$ and
$\overline{\nu}_{HBO}=\case{1}{2}(\nu_{HBO}+\case{1}{2}\nu_{2HBO})$. The
second value is a mean of $\nu_0$, calculated using the main resonance
frequency $\nu_{HBO}$ and the second harmonic frequency $\nu_{2HBO}$.
The average PDS and the main
frequencies $\nu_{HBO}$ (and also $\overline{\nu}_{HBO}$),
$\nu_1$ and $\nu_2$ were used
for the analysis of the TLM and RPM.
\subsection{Transition Layer Model (TLM)}
\begin{inlinefigure}
\centerline{\includegraphics[width=0.999\linewidth]{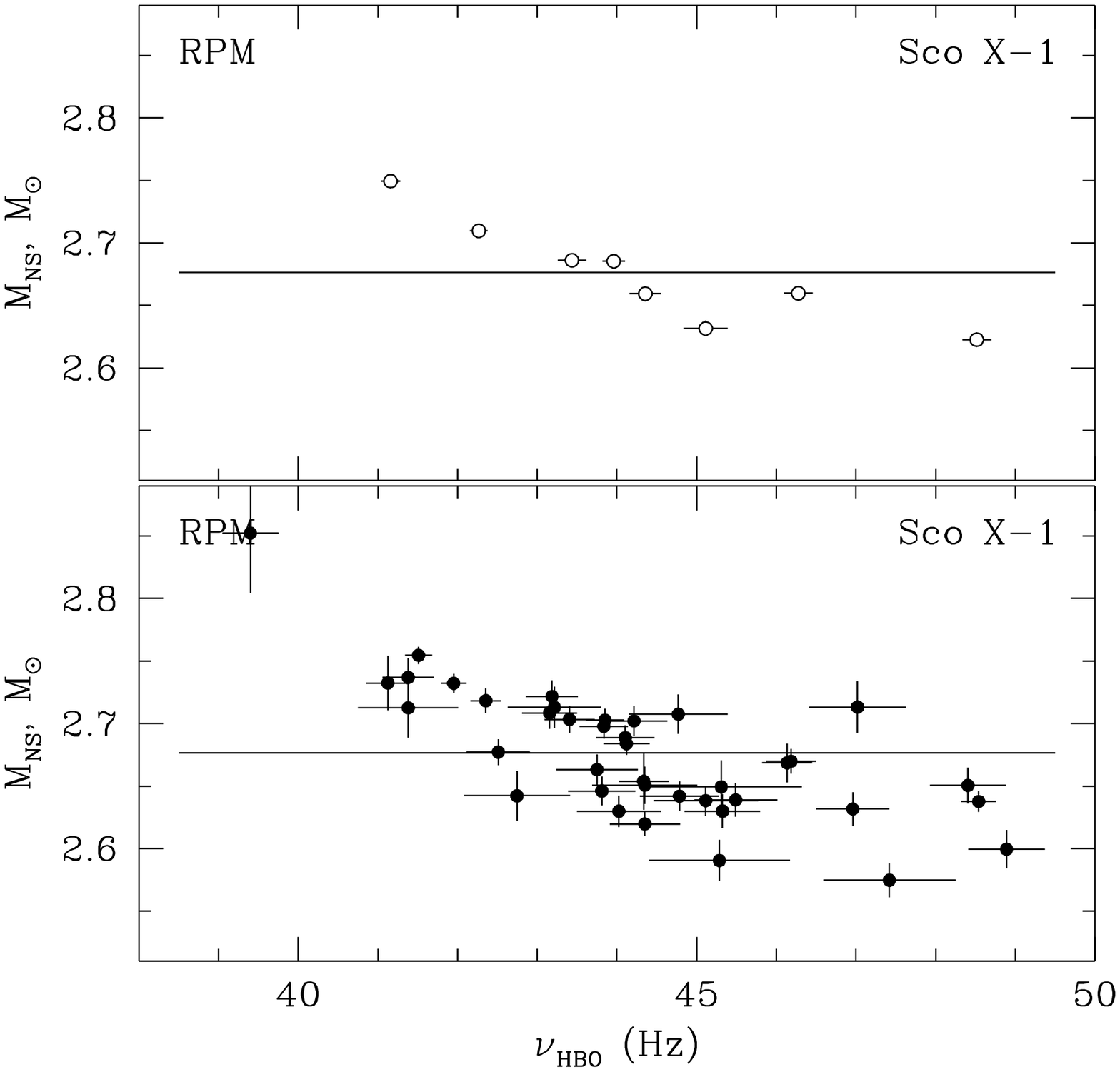}}
\caption{Dependence of the inferred neutron star mass, $M_{NS}$ on the
HBO frequency, $\nu_{HBO}$. The used data are the same as for Fig.~1
Approximations by a constant value are shown by solid lines}
\label{fig3} 
\end{inlinefigure}
For each set of $\nu_{HBO}$, $\nu_1$ and $\nu_2$ we obtained
the $\delta-$value using formula (1) in TO01, where $\nu_{HBO}$, $\nu_1$
and $\nu_2$ are identified with $\nu_L$, $\nu_{\rm K}$ and $\nu_h$
respectively.
In Figure~1 the dependence of 
the angle $\delta$ between a normal to the disk and a magnetospheric axis 
on $\nu_{HBO}$ is shown. 
We establish that the strongest dependence of $\delta$ on the QPO peak
frequencies involves $\nu_{HBO}$. Thus we investigated this dependence
only. The model invariants should be kept the same, independent of the
QPO frequency branch where it is verified.  In Table~1 the
approximation of $\delta$ by constant are present for two cases:
$\nu_L=\nu_{HBO}$ (see Fig.~1 the lower panel) 
and $\nu_L=\overline{\nu}_{HBO}=\case{1}{2}(\nu_{HBO}+\case{1}{2}\nu_{2HBO})$
(in the latter case the second harmonic is taken into account
for $\delta$ determination). For each case we provide a reduced
$\chi_{red}^2$. 
The $\delta-$rms and the standard deviations $\sigma_{\delta}$ are
also present in Table~1. We repeated the same procedure for $\delta$
vs $\nu_L$ when $\nu_L=\overline{\nu}_{HBO}$.  In Figure~1 these
results are present. In order to reduce the possible stochastic
scattering in the data, which always dominates relative to the
statistical scattering, we implement an averaging of $\delta$ for a
given HBO frequency (see the upper panel of Fig.~1).
We repeated the same procedure for $\delta$ vs $\nu_L$ when
$\nu_L=\overline{\nu}_{HBO}$. In Figure~2 these results are present.
Whereas some correlation of $\delta$ vs $\nu_{HBO}$ are clearly seen
in Figure~1, points $(\delta, ~\overline{\nu}_{HBO})$ are almost scattered
along the horizontal line in Figure~2 (see the upper panel).

\subsection{Relativistic Precession model (RPM)} 
Now we investigate the RPM invariant, the NS mass $M_{NS}$ as a
function of $\nu_{HBO}=\nu_{nod}$ or as a function of
$\overline{\nu}_{HBO}=\nu_{nod}$.  The deduced dependences of $M_{NS}$
on the HBO frequencies are stronger than for the kHz frequencies
(similar to the case with $\delta$ for the TLM).  In Table~2 and in
Figs.~3--4 we present a mean value of $M_{NS}$, $\chi^2_{red}$ and
$\sigma_{M}$. It is seen that $\chi^2_{red}$ values are a factor of
3-6 bigger than that in Table~1 for $\delta$. But the relative
deviation of $M_{NS}$, $\sigma_M$ is also 2\% as $\sigma_{\delta}$ for
$\delta$.
The linear correlation of the $M_{NS}$ values on HBO frequencies
$\nu_{HBO}$ is clearly seen in Fig.~3 and gets even stronger for
$\overline{\nu}_{HBO}$ (Fig.~4).
\begin{inlinefigure}
\centerline{\includegraphics[width=0.999\linewidth]{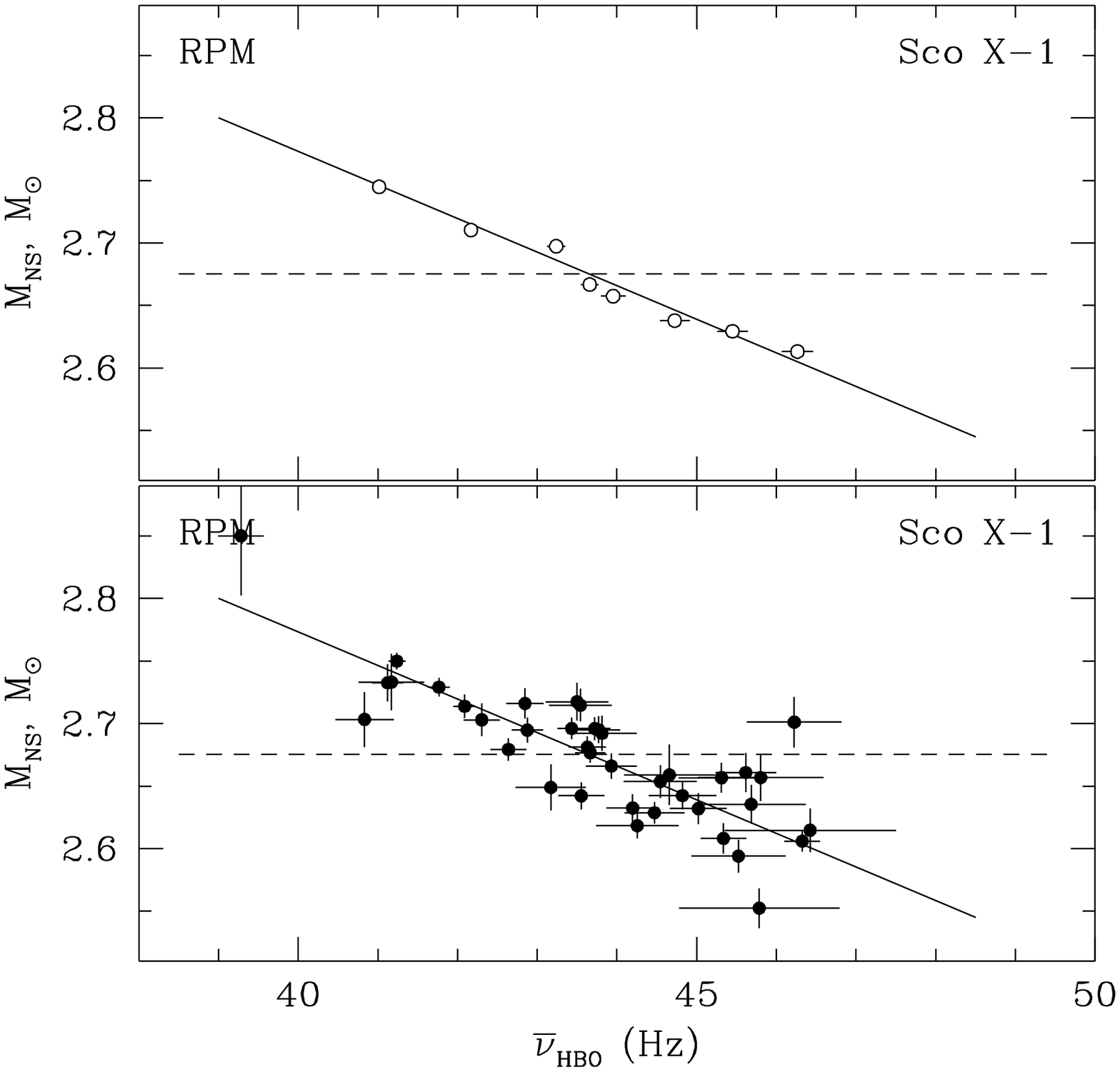}}
\caption{Dependence of the inferred neutron star mass on the HBO frequency, 
$\overline{\nu}_{HBO}$. The used data are the same as for Fig.~2.
Approximations by a constant and linear trend are shown by dashed and
solid lines respectively}
\label{fig4} 
\end{inlinefigure}
$M_{NS}$ drops from 2.8 $M_{\odot}$ to 2.6 $M_{\odot}$ with
increase of $\overline{\nu}_{HBO}$ from 39.5 Hz to 46.5 Hz\footnote{As it is
expected from the probability theory, the statistical mean $\overline
{M}_{NS}$ and the statistical deviation converge to the real values of
the mean and the standard deviation with increase of a number of
points for a given HBO frequency.}.  In RPM the mean $M_{NS}$ are
scattered along the oblique line whereas in the TLM $\delta$ are
scattered along the horizontal line.
\begin{inlinetable}
\caption{}
\footnotesize
\begin{center}
\begin{tabular}{ccc}
\hline
\hline
Parameter & $\nu_{L}=\nu_{HBO}$ & $\nu_{L}=\bar{\nu}_{HBO}$ \\
\hline
$\delta$   & 5.612$\pm$0.007     & 5.545$\pm$0.006\\
$\chi^{2}_{red}$ & 5.53          & 2.94\\
\mbox{$\overline{\delta}$} & 5.600$\pm$0.114 & 5.555$\pm$0.086  \\
\hline
$r_{s}(\delta)$ & 0.567& 0.359\\
$Prob$ &  2$\cdot$10$^{-4}$ & 3$\cdot$10$^{-2}$\\
$N$, bins & 39 & 39\\
\hline
$r_{s}(\delta)$ & 0.738 & 0.595 \\
$Prob$ &  3.7$\cdot$10$^{-2}$& 1.2$\cdot$10$^{-1}$\\
$N$, bins & 8 & 8\\
\hline
\end{tabular}
\end{center}
\aftertabcaption{See caption for Figs. 1-2: $\overline{\nu}_{HBO}=\case{1}{2}
(\nu_{HBO} + \case{1}{2}\nu_{2HBO})$, $\delta$ is the best--fit 
parameter, $\overline{\delta}=\case{1}{N}\sum{\delta_i}.$}
\end{inlinetable}

\subsection{Statistical evaluation  of  the suggested invariants for the TLM and the RPM}
In order to check the consistency of the TLM and the RPM we use the
Spearman's non-parametric test for evaluation of the correlation of
$\delta$ and $M_{NS}$ with HBO frequencies. If the correlation
coefficients $r_s=\pm1$, then there is a perfect linear correlation
between two values. In the general case $|r_s|\leq1$.
For each case we
also calculate a probability $Prob$ that the derived correlation is a
result of the statistical fluctuations \footnote{Particularly, this test is
useful for comparison whether one of the correlations is stronger than
another.}. 
\begin{inlinetable}
\caption{}
\footnotesize
\begin{center}
\begin{tabular}{ccc}
\hline
\hline
Parameter & $\nu_{nod}=\nu_{HBO}$ & $\nu_{nod}=\overline{\nu}_{HBO}$\\
\hline
$M_{NS}, M_{\odot}$ & 2.680$\pm$0.002 & 2.675$\pm$0.002\\
$\chi^{2}_{red}$ & 14.3  & 16.3\\
$\overline{M}_{NS}, M_{\odot}$ & 2.677$\pm$0.050 & 2.674$\pm$0.053\\
\hline
$a$, cm $\times10^5$ & 1.367$\pm$ 0.002 & 1.358$\pm$0.002\\ 
$\chi^{2}_{red}$ & 8.2 & 13.7\\
\hline
$N$, bins & 39 & 39\\
$r_{s}(M_{NS})$ & -0.715 & -0.789\\
$Prob$ &  3$\cdot$10$^{-7}$ & 2$\cdot$10$^{-9}$ \\
\hline
$r_{s}(M_{NS})$ &  -0.929 & -1.000\\
$Prob$ &  8.6$\cdot$10$^{-4}$& 0.0\\
$N$, bins & 8 & 8\\
\hline
\end{tabular}
\end{center}
\aftertabcaption{See caption for  Figs. 3-4: $a=J/(c\cdot M_{NS})$,
$\overline{\nu}_{HBO}=\case{1}{2} (\nu_{HBO} + \case{1}{2}\nu_{2HBO})$,
${M}_{NS}$ is best--fit parameter,
$\overline{M}_{NS}=\case{1}{N}\sum{M_{NS,i}}$.}
\end{inlinetable}

From Table~2 it is seen that the RPM is inconsistent with data
($r_s=-1$, {\it there is a perfect linear correlation between $M_{NS}$
and $\overline{\nu}_{HBO}$ ) and the proposed parameter $M_{NS}$ is
not invariant with respect to the observed QPO frequencies}.  It is
worth noting that inclusion of the second HBO component into the
determination of the HBO peak leads to a weakening of the correlation
for $\delta$ vs HBO frequencies while the opposite effect is seen for
the correlation between $M_{NS}$ vs HBO frequencies. In Figs.~2,~4 it
is shown that the averaging leads to a decrease in the scattering of
the inferred parameters along the best-fit line.  For the TLM (see
Fig.~2) $\delta$ is practically a constant, $\bar\delta=5\fdg55$ with
$\sigma_{\delta}\approx 0.5\%$ while for the RPM (Fig.~4) $M_{NS}$
strongly correlates with $\overline{\nu}_{HBO}$ and is inconsistent
with a constant.  It should be pointed out that $\chi^2_{red}$ in the
Table~1 are affected by the aforementioned underestimation of the
systematic and random resonance shift (see Introduction and \S 2.2).
Thus, the $\chi^2_{red}$ is artificially high because it takes into
account only statistical errors, and ignores this effect.

\section{Conclusions} 

Using the thorough data analysis of RXTE data for Sco X-1 we have
presented a detailed investigation of the transition layer (TL) and
the relativistic precession (RP) models. These models predict the
existence of the invariants: $\delta$ for the TLM and $M_{NS}$ for the
RPM.  We establish that: (1) The inferred $\delta-$value is
$5\fdg56\pm0\fdg09$ is consistent with a constant. Correlation of the
$\delta-$values with the HBO frequency is rather weak.  (2) There is a
strong correlation between an inferred neutron star mass and HBO
oscillation frequency in the RPM frameworks, (3) The inferred NS mass
$M_{NS}=(2.7\pm 0.1)M_{\odot}$ cannot be consistent
with any EOS of NS matter.

We are grateful the referee for the useful suggestions and the
valuable corrections. L.T. acknowledges the fruitful discussions with
Chris Shrader, Charlie Bradshaw, Vladimir Krasnopolski and Paul Ray.


\begin{thebibliography}{} 
\bibitem[Bardeen et al. (1972)]{bard}
Bardeen, J.M., Press, W.H., Teukolski, S.A. 1972, ApJ, 178, 374

\bibitem[Jahoda et al. (1996)]{jah} Jahoda, et. al. 1996, Proc. SPIE,
2808, 59

\bibitem[Jonker et al. (2002)]{jon}
Jonker, P.J., et al. 2002, MNRAS, in press (astro-ph/0202420)

\bibitem[Kato (1990)]{kato}
Kato, S. 1990, PASJ, 42, 99

\bibitem[Landau \& Lifshitz (1965)]{LL65} 
Landau, L.D. \& Lifshitz, E.M. 1965, {\it Mechanics}, New York: Pergamon Press
(LL)

\bibitem[Miller et al. (1998)]{mil}
Miller, M.C., Lamb, F.K., Psaltis, D. 1998, ApJ 508, 791.

\bibitem[Okazaki et al. (1987)]{okaz87}
Okazaki, T., Kato, S., \& Fukue, J. 1987, PASJ, 39, 457

\bibitem[Osherovich \& Titarchuk (1999)]{ot99}
Osherovich, V., \& Titarchuk, L. 1999, ApJ, 522, L113 (OT99)

\bibitem[Stella et al. (1999)]{stel}
Stella, G., Vietri, M., \& Morsink, S.M. 1999, ApJ, 524, L66

\bibitem[Strohmayer et al. (1996)]{str}
Strohmayer, T.E., et al.  1996, ApJ, 469, L9

\bibitem[Titarchuk, Lapidus \& Muslimov (1998)]{tlm}
Titarchuk, L.G., Lapidus, I.I.   \& Muslimov, A.  1998, \apj, 499, 315 (TLM98)

\bibitem[Titarchuk, \& Osherovich  (2001)]{to01}
Titarchuk, L., \& Osherovich, V.  2001, \apj, 555, L55

\bibitem[van der Klis (2000)]{vdk}
van der Klis, M. 2000, ARA\&A, 38, 717
 
\bibitem[van der Klis et al. (1996)]{vdk96}
van der Klis,  et al. 1996, \apj, 469, L1

\bibitem[Vikhlinin et al. (1994)]{vihkl}
Vikhlinin, A., Churazov, E., \& Gilfanov M. 1994, A\&A, 287, 73
\end{thebibliography}
\end{document}